\documentclass[conference]{IEEEtran}

\usepackage{graphicx,times,amsmath} 
\usepackage{amsmath}
\usepackage{amssymb}
\usepackage{amsfonts}
\usepackage{amsthm}
\usepackage{multicol}

\newtheorem*{stmt}{\sf Statement}

\newtheorem*{cor}{\sf Corrolary}
\theoremstyle{definition}
\newtheorem*{defi}{\sf Definition}
\newtheorem*{remark}{\sf Remark}

\IEEEoverridecommandlockouts  
\begin{document}

\title{\ \\ \LARGE\bf Probabilistic Dynamic Logic of Phenomena and Cognition
\thanks{Evgenii Vityaev is with the Department of Mathematical Logic,
Sobolev Institute of Mathematics of the Russian Academy of
Sciences and with the Department of Discrete mathematics and
Informatics of the Novosibirsk State University, 630090,
Novosibirsk, Russia, email: vityaev@math.nsc.ru}
\thanks{Boris Kovalerchuk is with the Department of Computer Science,
Central Washington University, Ellensburg, WA 98926-7520, e-mail:
borisk@cwu.edu}
\thanks{Leonid Perlovsky is with the Harvard University and the Air Force Research Laboratory, Sensors
Directorate, Hanscom AFB, leonid@seas.harvard.edu}
\thanks{Stanislav Smerdov, Novosibirsk State University, Sobolev Institute of Mathematics of the Russian Academy of
Sciences, 630090, Novosibirsk, Russia, email: netid@ya.ru} }

\author{Evgenii Vityaev, Boris Kovalerchuk, Leonid Perlovsky, Stanislav Smerdov}

\maketitle

\begin{abstract}
The purpose of this paper is to develop further the main concepts
of Phenomena Dynamic Logic (P-DL) and Cognitive Dynamic Logic
(C-DL), presented in the previous paper. The specific character of
these logics is in matching vagueness or fuzziness of similarity
measures to the uncertainty of models. These logics are based on
the following fundamental notions: \emph{generality relation,
uncertainty relation, simplicity relation, similarity maximization
problem with empirical content and enhancement (learning)
operator}. We develop these notions in terms of logic and
probability and developed a Probabilistic Dynamic Logic of
Phenomena and Cognition (P-DL-PC) that relates to the scope of
probabilistic models of brain. In our research the effectiveness
of suggested formalization is demonstrated by approximation of the
expert model of breast cancer diagnostic decisions. The P-DL-PC
logic was previously successfully applied to solving many
practical tasks and also for modelling of some cognitive
processes.
\end{abstract}

\section{Introduction}

In the paper \cite{Perlovsky&Kovelerchuk-DLPC:08} there was
introduced a Phenomena Dynamic Logic (P-DL) and Cognitive Dynamic
Logic (C-DL) as a generalization of the Dynamic Logic and Neural
Modelling Fields theory (NMF) introduced in the previous papers
\cite{Perlovsky, Perlovsky2}. Logics P-DL, C-DL provide the most
general description of Dynamic Logic in the following fundamental
notions\emph{generality relation, uncertainty relation, simplicity
relation, similarity maximization problem with empirical content
and enhancement (learning) operator}. This generalization provide
interpretation of P-DL, C-DL logics in the frame of other
approaches.

\noindent In this paper we interpret logics P-DL, C-DL in terms of
logic and probability: uncertainty we interpret as probability,
while the process of learning as a semantic probabilistic
inference \cite{Vityaev-asian:06, Vityaev-book:06,
Smerdov&Vityaev-CI:09, Smerdov&Vityaev-semr:09}. We also interpret
mentioned fundamental notions. The resulting Probabilistic Dynamic
Logic of Phenomena and Cognition (P-DL-PC) belong to the scope of
the probabilistic models of brain \cite{ProbMind, ProbMod}. Thus,
through logics P-DL, C-DL we extend the interpretation of Dynamic
Logic and Neural Modelling Fields theory to probabilistic models
of brain. The P-DL-PC logic as probabilistic model of brain was
previously applied to modelling of some cognitive process
\cite{Principals, Animat, Vityaev-book:06, SciDis}. \noindent The
effectiveness of P-DL-PC logic demonstrated in this paper by
approximation of the expert model of breast cancer diagnostic
decisions.

\section{Universal productions. Data for prediction}

\noindent In our study learning models will be generated as sets
of \emph{universal productions} (\emph{u-productions}), which are
introduced in this section. Note that every set of \emph{universal
formulas} is logically equivalent to a certain set of
u-productions.

\smallskip \noindent Consider a fixed first-order
language $\frak{L}$ in a countable signature. Hereafter denote
$\mathbf{A}_\frak{L}$ the set of all atoms; $\mathbf{L}_\frak{L}$
-- the set of all literals; $\mathbf{S}_\frak{L}^0$ -- the set of
ground sentences. The set of ground atoms and the set of ground
literals are denoted $\mathbf{A}_\frak{L}^0 \rightleftharpoons
\mathbf{A}_\frak{L} \cap \mathbf{S}_\frak{L}^0$ and
$\mathbf{L}_\frak{L}^0 \rightleftharpoons \mathbf{L}_\frak{L} \cap
\mathbf{S}_\frak{L}^0$ correspondingly. Following examples of
atoms and literals are given in the section VIII for the task of
approximation of the expert model of breast cancer diagnostic
decisions: `number of calcifications per $cm^3$ less than 20`,
`volume of calcifications in~$cm^3$ not less or equal to 5`,
`total number of calcifications more than 30 and etc.

\noindent Let $\Theta$ be the set of all substitutions and
$\Theta^0 \subseteq \Theta$ the set of ground substitutions, that
are mappings variables to ground terms. All necessary notions from
model theory and logic programming are elementary and can be
easily found in books \cite{Keisler&Chang-MT:90},
\cite{Maltsev-AS:70, Lloyd-LP:87}.

\begin{defi} A record of the type
$$\mathrm{R} \leftrightharpoons \tilde{\forall}
\left( \mathrm{A}_1 \wedge \cdots \wedge \mathrm{A}_m  \Leftarrow \mathrm{B}_1 \wedge
\cdots \wedge \mathrm{B}_n \right),$$ where $\mathrm{A}_1, \cdots \mathrm{A}_m,
\mathrm{B}_1, \cdots ,\mathrm{B}_n $ are literals, and $\tilde{\forall}$ stands for a bloc
of quantifiers over all free variables of the formulae in brackets (universal closure), is
called a \emph{u-production}. \emph{A variant of u-production $\mathrm{R}$} is
$$\mathrm{R} \theta
\rightleftharpoons \tilde{\forall} \left( \mathrm{A}_1 \theta
\wedge \cdots \mathrm{A}_m \theta \Leftarrow \mathrm{B}_1 \theta
\wedge \cdots \wedge \mathrm{B}_n \theta \right),$$ where $\theta$
is an arbitrary one-to-one correspondence over the set of
variables. Let ${\tt Prod}$ be the set of all u-productions.
\end{defi}

\noindent For example in section X presented the following
u-production that was discovered by the learning model:

\begin{quote}
\textsc{IF} TOTAL number of calcifications is more than 30,
\textsc{and} VOLUME is more than 5 $cm^3$,
\textsc{and} DENSITY of calcifications is moderate, \\
\textsc{THEN} Malignant.
\end{quote}

\noindent Let $\mathtt{Fact}_v \subset \mathbf{A}_\frak{L}$ be a
set of atoms from A that are valid for verification in algebraic
system ${\mathfrak{B}}$ appearing in practice. Our aim is to
investigate as much ``extra'' facts about $\frak{L}$ as possible,
i.e., to predict or explain them. A natural assumption is that we
can verify (falsify) each element of
\[ \mathtt{Fact}_{\rm o} \leftrightharpoons \left\{ {\rm A} \theta
\mid \theta \in \Theta^0,~{\rm A} \in \mathtt{Fact}_v \right\}.
\]

Certainly we may postulate our ability to check any literal of
$\mathtt{Fact}_v^\ast \leftrightharpoons \mathtt{Fact}_v \cup
\left\{ \neg {\rm A} \mid {\rm A} \in \mathtt{Fact}_v \right\}$.
For the rest of the literals (and their conjunctions) the
machinery of \emph{probabilistic prediction} will be defined later
on. Note that ${\tt Fact}_{\rm o}^\ast = {\tt Fact}_{\rm o} \cup
{\tt Fact}_{\rm o}^\neg$, where ${\tt Fact}_{\rm o}^\neg
\rightleftharpoons \left\{ \neg \mathrm{A} \mid \mathrm{A} \in
{\tt Fact}_{\rm o} \right\}$ is \emph{the complete set of
alternatives} allowing a real test.

The \emph{data} are defined as a maximal (logically) consistent
subset of the complete set of alternatives, i.e., being given a
mapping \mbox{$\zeta _\mathfrak{B} :{\tt Fact}_{\rm o} \mapsto
\left\{ {\bot,\top} \right\}$} (here \emph{$\bot$ -- ``false'',
$\top$ -- ``true''}) we conclude that $${\tt Data}\left[
\mathfrak{B} \right] \rightleftharpoons \left\{ {{\rm A} \mid {\rm
A}\in {\tt Fact}_{\rm o}~\mbox{\textit{and}}~ \zeta _\mathfrak{B}
\left( {\rm A} \right)= \top} \right\}
 \cup $$ $$ \left\{ {\neg{\rm A} \mid {\rm A}\in {\tt
Fact}_{\rm o}~\mbox{\textit{and}}~ \zeta _\mathfrak{B} \left( {\rm A} \right)=\bot}
\right\}.$$


\section{Generality relation between theories}

\noindent The idea of a \emph{generality relation} between
theories can be viewed, for example, as a reduction of the set of
properties predicted by the use of these theories. A more general
theory (potentially) predicts a greater number of formal features.
We start with a generality relation between one-element
specifications, i.e., between u-productions.

\begin{defi} For two productions $\mathrm{R}_1 \equiv \tilde{\forall} \left(\mathrm{A}_1
\wedge \cdots \wedge \mathrm{A}_{m_1} \Leftarrow \mathrm{B}_1 \wedge
\cdots \wedge \mathrm{B}_{n_1} \right) $ and $\mathrm{R}_2 \equiv
\tilde{\forall} \left(\mathrm{C}_1 \wedge \cdots \wedge
\mathrm{C}_{m_2} \Leftarrow \mathrm{D}_1 \wedge \cdots \wedge
\mathrm{D}_{n_2} \right) $ a relation $\mathrm{R}_1 \succ
\mathrm{R}_2 $ (\textit{``more general than''}) takes place if and
only if there exists $\theta \in \Theta $ such that $\left\{
{\mathrm{B}_1 \theta ,\cdots,\mathrm{B}_{n_1} \theta } \right\}
\subseteq \left\{ {\mathrm{D}_1 ,\cdots, \mathrm{D}_{n_2 } }
\right\}$, $\left\{ {\mathrm{A}_1 \theta ,\cdots,\mathrm{A}_{m_1}
\theta } \right\} \supseteq \left\{ {\mathrm{C}_1 ,\cdots,
\mathrm{C}_{m_2 } } \right\}$, and $n_1 \leqslant n_2$, $m_1
\geqslant m_2$, $\not \vdash {\rm R}_1 \equiv {\rm R}_2$.
\end{defi}

\noindent The inclusion of the sets of premises designates that
the more general u-production is, then the wider its field of
application. The inverse inclusion (for conclusions) says that
$\mathrm{R}_1$ predicts a greater number of properties using a
smaller premise.

Let $S \subseteq {\tt Prod}$. Denote ${\tt Fact} \left[ S ;
\frak{B} \right]$ the set of all $\mathrm{A} \in
\mathbf{L}_\frak{L}^0$ such that for some ${\rm R} \in S$ and
$\theta \in \Theta^0$,
\begin{center}
$\mathrm{R} \theta \equiv \left(\mathrm{A}_1 \wedge \cdots \wedge
\mathrm{A}_{m} \Leftarrow \mathrm{B}_1 \wedge \cdots \wedge
\mathrm{B}_{n} \right)$, holds \\ $\left\{ \mathrm{B}_1 , \cdots ,
\mathrm{B}_{n} \right\} \subseteq {\tt Data} \left[ \frak{B}
\right]$ and $\mathrm{A} \in \left\{ \mathrm{A}_1 , \cdots ,
\mathrm{A}_{m} \right\}$.
\end{center}

\noindent Thus, ${\tt Fact} \left[ S ; \frak{B} \right]$ is the set
of ground literals predicted according to available data (about the
model $\frak{B}$) together with u-productions in $S$.

\smallskip
\noindent In the sequel let $\succcurlyeq$ be a reflexive closure of
$\succ$. One should pay attention to the fact: $\mathrm{R}_1
\succcurlyeq \mathrm{R}_2 $ entails that ${\tt Fact} \left[ \left\{
\mathrm{R}_1 \right\} ; \frak{B} \right]$ contains ${\tt Fact}
\left[ \left\{ \mathrm{R}_2 \right\} ; \frak{B} \right]$.

\medskip \noindent Thereafter it isn't difficult to extend the domain
of our generality relation to subsets of ${\tt Prod}$.

\begin{defi} Let $S,S' \subseteq {\tt Prod}$, and for
any ${\rm R}' \in S'$ we find ${\rm R} \in S$ such that $\mathrm{R} \succcurlyeq
\mathrm{R}'$. In this case we say `$S$ is not less general than $S'$' ($S \vartriangleright
S'$).
\end{defi}

\noindent It's straightforward to notice that ${\tt Fact} \left[
S' ; \frak{B} \right] \subseteq {\tt Fact} \left[ S ; \frak{B}
\right]$ for $S$ and $S'$ from the definition above. Remark that
$S$ may include u-productions apart from those, which are
generalizations of elements of $S'$.


\section{Probability/degree of belief}

\noindent The topic of distributing probability over formulas of
propositional logic (as well as over ground statements in a first
order language) being widely discussed in a literature and meets
Kolmogorov's understanding of probability measure \cite{Shiryaev-P}.
The following definition is given on the basis of analysis cited
in~\cite{Halpern:90}.

\begin{defi} \textit{A probability} over $F\subseteq {\mathbf S}_\mathfrak{L}^0 $
closed with respect to $\wedge$, $\vee $ and $\neg $, is a function
\mbox{$\mu:F\mapsto \left[{0,1}\right]$} satisfying the following
conditions:

\begin{enumerate}
\item if $\vdash \Phi $ (``$\Phi $
is a tautology''), then $\mu \left( \Phi \right)=1$;

\item if $\vdash \neg \left( {\Phi \wedge \Psi } \right)$, then $\mu \left( {\Phi \vee \Psi
} \right)=\mu \left( \Phi \right)+\mu \left( \Psi \right)$. \end{enumerate}

\end{defi}

For any ground instance of a u-production its probability is
defined as conditional, i.e.,
\begin{center}
$\mu \left( {\rm A}_1 \wedge \cdots {\rm A}_m \Leftarrow {\rm B}_1
\wedge \cdots\wedge {\rm B}_n \right) =$ \\ $= \mu \left( {\rm A}_1
\wedge \cdots {\rm A}_m \mid {\rm B}_1 \wedge \cdots\wedge {\rm B}_n
\right) = \frac{\mu \left( {{\rm A}_1 \wedge \cdots {\rm A}_m \wedge
{\rm B}_1 \wedge \cdots\wedge {\rm B}_n } \right)}{\mu \left( {{\rm
B}_1 \wedge \cdots\wedge {\rm B}_n } \right)}$
\end{center}

Let ${\rm R} \in {\tt Prod}$. Denote as ${\tt Sub} \left[ {\rm R}
\right]^\mu$ those substitutions~$\theta \in \Theta^0$, for which
the premise of u-production ${\rm R} \theta$ has a non-zero
probability.

\begin{center}
${\tt Prod}^\mu \rightleftharpoons \left\{ {\rm R}\in {\tt Prod}
\mid {\tt Sub} \left[ {\rm R} \right]^\mu \ne \varnothing \right\}$;

\smallskip $\underline{\mu }\left( {\rm R} \right)\rightleftharpoons
{\rm inf}\left\{ \mu \left( {{\rm R}\theta } \right) \mid \theta \in
{\tt Sub} \left[ {\rm R} \right]^\mu \right\}$, where ${\rm R} \in
{\tt Prod}^\mu$.
\end{center}

\noindent A value of conditional probability serves to
characterize our \emph{degree of belief} (and responsible for an
\emph{uncertainty relation}) in reliability of different causal
connections included in temporary specification. Note that two
productions are not necessary comparable with respect to
generality relation $\succcurlyeq$; moreover, their premisses may
not be contained in the complete set of alternatives (and so these
productions will be not valid for a direct check in a real
structure $\mathfrak{B}$).


\section{Simplicity of probabilistic theories}

\noindent Adding comparison of lower probabilistic estimations to
the definition of generality relation we obtain the following
definition.

\begin{defi} Let $S,S' \subseteq {\tt Prod}^\mu$.
We say that $S$ is \emph{more $\mu$-general than} $S'$ iff for every $\mathrm{C}' \in S'$
there exists $\mathrm{C} \in S$ such that $\mathrm{C} \succcurlyeq \mathrm{C}'$ and
$\underline{\mu }(\mathrm{C}) \geqslant \underline{\mu }(\mathrm{C}')$, and in at least one
of the cases the strong relation $\succ$ takes place.
\end{defi}

\medskip
\noindent Hence, $\mu$-generalization allows us to define a more
general set $S$ in such a way that the lower estimations of
probabilities is not declined. When our belief to the elements of
$S$ is no less than that of $S'$, then we have a \emph{simplicity
relation} -- the set $S$ is simpler than $S'$ in order to
describe/predict the properties.


\section{Similarity measure with the empirical content}

\noindent By elaboration of u-productions we mean the gain of its conditional probability.

\smallskip
\begin{defi} A relation ${\rm R}_1 \sqsubset {\rm R}_2 $ (\textit{`probabilistic inference'})
for ${\rm R}_1 ,{\rm R}_2 \in {\tt Prod}^\mu $ means that ${\rm R}_1 \succ {\rm R}_2 $ and $\underline{\mu }\left( {{\rm R}_1 } \right)<\underline{\mu }\left( {{\rm R}_2 }
\right)$.
\end{defi}

\begin{defi} Let $\pi$ be some requirements to be applied to elements of ${\tt Prod}^\mu$, i.e.
$\pi: {\tt Prod}^\mu \mapsto \left\{ \bot, \top \right\}$ (value
is equal to $\bot$, if u-production satisfies $\pi$, and $\top$ --
otherwise); $\mathsf{\Pi} \leftrightharpoons \left\{ {\rm R} \in
{\tt Prod}^\mu \mid \pi \left( {\rm R} \right) = \top \right\}$.
We say that $\mathrm{R}_2 \in \mathsf{\Pi}$ is a \emph{minimal
follower of $\mathrm{R}_1 \in {\tt Prod}^\mu$ relative to
$\sqsubset$ in $\mathsf{\Pi}$} (denoted as $\mathrm{R}_1
\sqsubset_\pi \mathrm{R}_2$), iff $\mathrm{R}_1 \sqsubset
\mathrm{R}_2$ and there is no intermediate u-production
$\mathrm{R}_{3/2} \in \Pi$ such that $\mathrm{R}_1 \sqsubset
\mathrm{R}_{3/2} \sqsubset \mathrm{R}_2$.
\end{defi}

\noindent In the prediction of a literal ${\rm H}$ the \emph{similarity measure} for
u-productions, which are valid for verification and applicable to the goal ${\rm H}$,
is equal to conditional probability $\underline{\mu }\left( \cdot \right)$.
Thus we deal with a uniform measure of similarity.


\section{Learning operator}

\begin{defi} A production \[\mathrm{R} \equiv
\tilde{\forall} \left( \mathrm{A}_1 \wedge \cdots \wedge
\mathrm{A}_m \leftarrow \mathrm{B}_1 \wedge \dots \wedge
\mathrm{B}_n \right) \] is called a \emph{maximal specific
u-production (ums-production) for prediction of a conjunction
$\mathrm{H} \equiv \left( \mathrm{H}_1 \wedge \cdots \wedge
\mathrm{H}_k \right)$,} where $\left\{ \mathrm{H}_1 ,\cdots
,\mathrm{H}_k \right\} \subset \mathbf{L}_\frak{L}$ and $m
\leqslant k$, iff the following conditions are satisfied:

\begin{enumerate}
\item there is a substitution $\theta$ (not necessary ground) such that $\left\{ \mathrm{A}_1
,\cdots ,\mathrm{A}_m \right\} \subseteq \left\{ \mathrm{H}_1 \theta ,\cdots ,\mathrm{H}_k
\theta \right\}$, $\left\{ {\mathrm{B}_1 ,\dots,\mathrm{B}_n } \right\}\subseteq \left\{
\mathrm{B} \theta \mid \mathrm{B} \in \mathtt{Fact}_v^\ast \right\}$;

\item if $\mathrm{D} \in \left\{ \mathrm{A}_1 ,\cdots
,\mathrm{A}_m \right\}$ and $\theta_{\rm o} \in {\tt Sub} \left[
{\rm R} \right]^\mu$, then \\ $\mu \left( {\mathrm{A}_1
\theta_{\rm o} \wedge \cdots \wedge \mathrm{A}_m} \theta_{\rm o}
\right) < \\ \mu \left( {\mathrm{A}_1 \theta_{\rm o} \wedge \cdots
\wedge \mathrm{A}_m} \theta_{\rm o} \mid
{\mathrm{B}_1 \theta_{\rm o} \wedge \cdots \wedge \mathrm{B}_n} \theta_{\rm o} \right)$ \\
and $\mu \left( \mathrm{D} \theta_{\rm o} \right) < \mu \left( \mathrm{D} \theta_{\rm o}
\mid {\mathrm{B}_1 \theta_{\rm o} \wedge \cdots \wedge \mathrm{B}_n} \theta_{\rm o}
\right)$;

\item there is no ${\rm R'} \in \mathtt{Prod}^\mu$, for which
points (1--2) are hold along with ${\rm R} \sqsubset {\rm R'}$;

\item the u-production ${\rm R}$ can't be generalized up to some
${\rm R'} \in \mathtt{Prod}^\mu$ satisfying all the previous
points (1--3) without decreasing its estimation $\underline{\mu }
\left( \cdot \right)$.
\end{enumerate} \end{defi}

\noindent The conditions above (for corresponding ums-productions)
are denoted as `point.i', $1 \leqslant i \leqslant 4$.

\begin{remark} Though condition point.4 emphasizes the nature of definition,
but it isn't necessary for indication. Indeed, if $\mathrm{R}$ may
be generalized up to $\mathrm{R}'$ under preserving point.1--3,
then $\underline{\mu } \left( \mathrm{R} \right) \leqslant
\underline{\mu } \left( \mathrm{R}' \right)$ (otherwise we get
${\rm R'} \sqsubset {\rm R}$ -- that contradicts point.3 for
$\mathrm{R}$.
\end{remark}

Let $\pi \left( {\rm R} \right) = \top$ be fulfilled for ${\rm R}
\in \mathtt{Prod}^\mu$ iff conditions points.1--2 are satisfied
for ${\rm R}$ and ${\rm H}$ (the last one is fixed from this
moment); denote $\mathsf{\Pi} \leftrightharpoons \pi^{-1} \left(
\top \right)$.

Define the \emph{probabilistic fix-point operator} $\mathrm{T}_\pi: 2^{\mathtt{Prod}^\mu} \mapsto
2^{\mathtt{Prod}^\mu}$ as follows: for a set $S \subseteq \mathtt{Prod}^\mu$ it produces

\begin{center}
$S' \leftrightharpoons \left\{ \mathrm{R}' \mid \mathrm{R} \sqsubset_\pi \mathrm{R}' \
\mbox{\emph{for some}} \ \mathrm{R} \in S \right\} \cup $ \\ $\cup \left\{ \mathrm{R} \mid
\mathrm{R} \in S \cap \mathsf{\Pi} \ \mbox{\emph{and there is no}} \ \mathrm{R}' \
\mbox{\emph{such that}} \ \mathrm{R} \sqsubset_\pi \mathrm{R}' \right\}$.
\end{center}

\noindent Therefore the operator $\mathrm{T}_\pi: S \mapsto S'$
possess important properties:

\begin{enumerate}
\item the set $S'$ is always more precise than $S$ (relative to $\succcurlyeq$);

\item the conditional probabilities $\underline{\mu } \left( \cdot
\right)$ increase during the conversion to more particular cases
(and so fuzziness decreases);

\item the similarity measure with the empirical content becomes greater for at least one
u-production (in $S$) when the operator converts $S$ to $S'$ (if not $S=S'$, of course);
\end{enumerate}

\noindent As a result the operator $\mathrm{T}_\pi$ is the \emph{enhancement, or learning, operator} in the sense of \cite{Perlovsky&Kovelerchuk-DLPC:08}.

\begin{defi} A fix-point (f.p., for short) $S$ of $\mathrm{T}_\pi$ is \emph{optimal} iff
there is no other f.p. $S'$ of considered operator, which is more $\mu$-general than $S$. \end{defi}

\begin{stmt} A subset $S \subseteq \mathtt{Prod}^\mu$ is a fix-point of the
operator $\mathrm{T}_\pi$ iff every element of $S$ satisfies
points.1--3 for $\mathrm{H}$.
\end{stmt}

\begin{cor} A subset $S \subseteq \mathtt{Prod}^\mu$ is an optimal fix-point of the operator $\mathrm{T}_\pi$ iff every element of $S$ is a ums-production for prediction of $\mathrm{H}$.
\end{cor}

\noindent Ums-productions may be viewed as a result of performing
generalized scheme of the \emph{semantic probabilistic inference}
\cite{Vityaev-asian:06, Smerdov&Vityaev-semr:09}, which is
realized by the fix-point operator described above. The program
system `Discovery' (see \cite{Vityaev&Kovalerchuk-finance:00,
KVR-Medicine, Vityaev-book:06, SciDis}) was developed: it carries
out the propositional version of the probabilistic fix-point
(learning) operator and was successfully applied to solving many
practical tasks \cite{SciDis}.


\section{Extraction of the expert model of breast cancer diagnostic decisions}

We applied our method to approximation of the expert model of breast cancer diagnostic
decisions that was obtained from the radiologist J.Ruiz \cite{KVR-Medicine}. At first we
extract this model from the expert by the special procedure using monotone boolean
functions \cite{KVR-Medicine} and then apply the program system `Discovery'
\cite{Vityaev&Kovalerchuk-finance:00} to approximate this model.

\subsection{Hierarchical Approach}
At first we ask an expert to describe particular cases using the
binary features. Then we ask a radiologist to evaluate a
particular cases, when features take on specific values. A typical
query will have the following format: "If feature 1 has value
$v_1$, feature 2 has value $v_2$, ..., feature n has value $v_n$,
then is a case suspicious of cancer or not?"

Each set of values ($v_1, v_2, ...,v_n$) represent a possible clinical case. It is
practically impossible to ask a radiologist to generate diagnosis for thousands of possible
cases. A hierarchical approach combined with the use of the property of monotonicity makes
the problem manageable. We construct a hierarchy of medically interpretable features from a
very generalized level to a less generalized level. This hierarchy follows from the
definition of the 11 medically oriented binary attributes. The medical expert indicate that
the original 11 binary attributes $w_1, w_2, w_3, y_1, y_2, y_3, y_4, y_5, x_3, x_4, x_5$
could be organized in terms of a hierarchy with development of two new generalized
attributes $x_1$, depending on attributes $w_1, w_2, w_3$, and $x_2$, depending on
attributes $y_1, y_2, y_3, y_4, y_5$.

A new generalized feature, $x_1$ -- `Amount and volume of calcifications' with grades (0 -
`benign' and 1 - `cancer') was introduced based on features: $w_1$ -- number of
calcifications/cm3, $w_2$ -- volume of calcification, cm3 and $w_3$ -- total number of
calcifications. We view $x_1$ as a function $g(w_1, w_2, w_3)$ to be identified. Similarly
a new feature $x_2$ -- `Shape and density of calcification' with grades:  (1) for `cancer'
and (0)-`benign' generalizes features: $y_1$ -- `irregularity in shape of individual
calcifications' $y_2$ --  `variation in shape of calcifications' $y_3$ -- `variation in
size of calcifications' $y_4$ --  `variation in density of calcifications' $y_5$ --
`density of calcifications'. We view $x_2$ as a function $x_2 = h(y_1, y_2, y_3, y_4, y_5)$
to be identified for cancer diagnosis.

As result we have a decomposition of our task as follows:
$$f \left(x_1, x_2, x_3, x_4, x_5 \right) = $$
$$ f \left( g \left(w_1, w_2, w_3 \right), h \left( y_1, y_2, y_3, y_4, y_5
\right), x_3, x_4, x_5 \right).$$


\subsection{Monotonicity}
Giving the above definitions we can represent clinical cases in
terms of binary vectors with five generalized features as:
$(x_1,x_2,x_3,x_4,x_5)$. Let us consider two clinical cases that
are represented by the two binary sequences: (10110) and (10100).
If radiologist correctly diagnose (10100) as cancer, then, by
utilizing the property of monotonicity, we can also conclude that
the clinical case (10110) should also be cancer. Medical expert
agreed with presupposition about monotonicity of the functions $f
\left( x_1, x_2, x_3, x_4, x_5 \right)$ and $h \left( y_1, y_2,
y_3, y_4, y_5 \right)$.

Let us describe the interview with an expert using minimal
sequence of questions to completely infer a diagnostic function
using monotonicity. This sequence is based on fundamental Hansel
lemma \cite{Hansel}. We omit a detailed description of the
specific mathematical steps. They can be found in \cite{KovTil}.
Table 1 illustrates this.

\subsection{Expert model extraction}
Columns 2 and 3 present values of above defined functions $f$ and
$h$. We omit a restoration of function $g \left( w_1, w_2, w_3
\right)$ because few questions are needed to restore this
function. All 32 possible cases with five binary features $\langle
x_1, x_2, x_3, x_4, x_5 \rangle$ are presented in column 1 in
table 1. They are grouped and the groups are called Hansel chains
\cite{KVR-Medicine}.  The sequence of chains begins with the
shortest chain 1 -- $(01100) < (11100)$ for five binary features.
Then largest chain 10 consists of 6 ordered cases: $(00000) <
(00001) <(00011) < (00111) < (01111) < (11111)$. The chains are
numbered there from 1 to 10 and each case has its number in the
chain, e.g., 1.2 means the second case in the first chain.
Asterisks in columns 2 and 3 mark answers obtained from an expert,
e.g., 1* for case (01100) in column 3 means that the expert
answered `yes'. The answers for some other chains in column 3 are
automatically obtained using monotonicity. The value f(01100) = 1
for case 1.1 is extended for cases 1.2, 6.3. and 7.3 in this way.
Similarly values of the monotone Boolean functions h are computed
using the table 1. The attributes in the sequence (10010) are
interpreted as $y_1, y_2, y_3, y_4, y_5$ for the function h
instead of $x_1, x_2, x_3, x_4, x_5$. The Hansel chains are the
same if the number of attributes is the same five in this case.

Column 5 and 6 list cases for extending functions' values without
asking an expert. Column 5 is for extending functions' values from
1 to 1 and column 6 is for extending them from 0 to 0. If an
expert gave an answer opposite (f(01100) = 0) to that presented in
table 1 for function $f$ in the case 1.1, then this 0 value could
be extended in column 2 for cases 7.1 (00100) and 8.1 (01000).
These cases are listed in column 5 for case (01100). There is no
need to ask an expert about cases 7.1 (00100) and 8.1 (01000).
Monotonicity provides the answer. The negative answer f(01100) = 0
can not be extended for f(11100). An expert should be queried
regarding f(11100). If his/her answer is negative f(11100) = 0
then this value can be extended for cases 5.1. and 3.1 listed in
column 5 for case 1.2. Relying on monotonicity, the value of f for
them will also be 0.

The total number of cases with asterisk (*) in columns 2 and 3 are equal to 13 and 12.
These numbers show that 13 questions are needed to restore the function $f \left( x_1, x_2,
x_3, x_4, x_5 \right)$ and 12 questions are needed to restore the function $h \left( y_1,
y_2, y_3, y_4, y_5 \right)$. This is only 37.5\% of 32 possible questions. The full number
of questions for the expert without monotonicity and hierarchy is $2^{11} = 2048$.


\section{Approximation of the expert model by learning operator}
For the Approximation of the expert model we used the program
system `Discovery' \cite{Vityaev&Kovalerchuk-finance:00}, that
realizes the propositional case of the probabilistic fix-point
learning operator. We discovered several dozens diagnostic rules
that were statistically significant on the 0.01, 0.05 and 0.1
levels of (F-criterion). Rules were extracted using 156 cases (73
malignant, 77 benign, 2 highly suspicious and 4 with mixed
diagnosis). In the Round-Robin test our rules diagnosed 134 cases
and refused to diagnose 22 cases. The total accuracy of diagnosis
is 86\%. Incorrect diagnoses were obtained in 19 cases (14\% of
diagnosed cases). The false-negative rate was 5.2\% (7 malignant
cases were diagnosed as benign) and the false-positive rate was
8.9\% (12 benign cases were diagnosed as malignant). Some of the
rules are shown in table 2. This table presents examples of
discovered rules with their statistical significance. In this
table:
\begin{itemize}
\item `NUM' -- number of calcifications per $cm^3$; \item `VOL' -- volume in~$cm^3$; \item
`TOT' -- total number of calcifications; \item `DEN' -- density of calcifications; \item
`VAR' -- variation in shape of calcifications; \item `SIZE' -- variation in size of
calcifications; \item `IRR' -- irregularity in shape of calcifications; \item `SHAPE' --
shape of calcifications.
\end{itemize}

We studied three levels of similarity measure: 0.7, 0.85 and 0.95.
A higher level of conditional probability decreases the number of
rules and diagnosed patients, but increases accuracy of diagnosis.


\onecolumn

\begin{center}
\begin{tabular}{|c|c|c|c|c|c|c|} \hline
\multicolumn{7}{|c|}{\textsf{\textbf{Table 1}.
Dynamic sequence of questions to expert}} \\
\hline \hline

{\tt 1} & {\tt 2} & {\tt 3} & {\tt 4} & {\tt 5} & {\tt 6} & {\tt 7} \\
\hline

    \multicolumn{1}{|c|}{\textsf{Number}}
    & \multicolumn{1}{|c|}{$f$}
    & \multicolumn{1}{|c|}{$h$}
    & \multicolumn{2}{|c|}{\textsf{Monotonic extrapolation}}
    & \multicolumn{1}{|c|}{\textsf{Chain}}
    & \multicolumn{1}{|c|}{\textsf{Case}}
    \\ \cline{4-5}

    & \multicolumn{1}{|c|}{\textsf{Diagnose}} & \multicolumn{1}{|c|}{\textsf{Form and V}}
    & \multicolumn{1}{|c|}{$1 \mapsto 1$}
    & \multicolumn{1}{|c|}{$0 \mapsto 0$}
    & & \\ \hline

$(01100)$ & 1* & 1* & 1.2, 6.3, 7.3 & 7.1, 8.1 & Chain 1 & 1.1 \\
\hline $(11100)$ & 1   & 1     & 6.4, 7.4     & 5.1, 3.1 &    & 1.2 \\
\hline $(01010)$ & 0* & 1* & 2.2, 6.3, 8.3 & 6.1, 8.1 & Chain 2 & 2.1 \\
\hline $(11010)$ & 1* & 1  & 6.4, 8.4     & 3.1, 6.1 &        & 2.2 \\
\hline $(11000)$ & 1* & 1* & 3.2         & 8.1, 9.1 & Chain 3 & 3.1 \\
\hline $(11001)$ & 1  & 1  & 7.4, 8.4     & 8.2, 9.2 &        & 3.2 \\
\hline $(10010)$ & 0* & 1* & 4.2, 9.3  & 6.1, 9.1 & Chain 4 & 4.1 \\
\hline $(10110)$ & 1* & 1  & 6.4, 9.4  & 6.2, 5.1 &        & 4.2 \\
\hline $(10100)$ & 1* & 1* & 5.2      & 7.1, 9.1 & Chain 5 & 5.1 \\
\hline $(10101)$ & 1  & 1  & 7.4, 9.4  & 7.2, 9.2 &        & 5.2 \\
\hline $(00010)$ & 0  & 0* & 6.2, 10.3 & 10.1    & Chain 6 & 6.1 \\
\hline $(00110)$ & 1* & 0* & 6.3, 10.4 & 7.1  &        & 6.2 \\
\hline $(01110)$ & 1  & 1  & 6.4, 10.5 &      &        & 6.3 \\
\hline $(11110)$ & 1  & 1  & 10.6     &      &        & 6.4 \\
\hline $(00100)$ & 1* & 0* & 7.2, 10.4 & 10.1 & Chain 7 & 7.1 \\
\hline $(00101)$ & 1  & 0* & 7.3, 10.4 & 10.2 &        & 7.2 \\
\hline $(01101)$ & 1  & 1* & 7.4, 10.5 & 8.2, 10.2 &    & 7.3 \\
\hline $(11101)$ & 1  & 1  & 5.6      &          &    & 7.4 \\
\hline $(01000)$ & 0  & 1* & 8.2      & 10.1 & Chain 8 & 8.1 \\
\hline $(01001)$ & 1* & 1  & 8.3      & 10.2 &        & 8.2 \\
\hline $(01011)$ & 1  & 1  & 8.4      & 10.3 &        & 8.3 \\
\hline $(11011)$ & 1  & 1  & 10.6     & 9.3  &        & 8.4 \\
\hline $(10000)$ & 0  & 1* & 9.2      & 10.1 & Chain 9 & 9.1 \\
\hline $(10001)$ & 1* & 1  & 9.3      & 10.2 &        & 9.2 \\
\hline $(10011)$ & 1  & 1  & 9.4      & 10.3 &        & 9.3 \\
\hline $(10111)$ & 1  & 1  & 10.6     & 10.4 &        & 9.4 \\
\hline $(00000)$ & 0  & 0  & 10.2     &      & Chain 10 & 10.1 \\
\hline $(00001)$ & 0* & 0  & 10.3     &      &         & 10.2 \\
\hline $(00011)$ & 1* & 0  & 10.4     &      &         & 10.3 \\
\hline $(00111)$ & 1  & 1* & 10.5     &      &         & 10.4 \\
\hline $(01111)$ & 1  & 1  & 10.6     &      &         & 10.5 \\
\hline $(11111)$ & 1  & 1  &          &      &         & 10.6 \\
\hline \textsf{Questions}  & 13 & 12  &      &         &          &      \\
\hline

\end{tabular}

\begin{tabular}{|c|c|c|c|c|c|c|} \hline

     \multicolumn{7}{|c|}{\textsf{\textbf{Table 2}. Examples of discovered diagnostic
     rules}} \\ \hline \hline

     \multicolumn{1}{|c|}{\textsf{Diagnosis}}
     & \multicolumn{2}{|c|}{\textsf{$f$-criteria}}
     & \multicolumn{3}{|c|}{\textsf{Value. $f$-criteria}}
     & \multicolumn{1}{|c|}{\textsf{Precision}}
     \\ \cline{4-6}

     \textsf{rule} & \multicolumn{2}{|c|}{} & \multicolumn{1}{|c|}{0.01}
     & \multicolumn{1}{|c|}{0.05}
     & \multicolumn{1}{|c|}{0.1} & \multicolumn{1}{|c|}{\textsf{on control}} \\ \hline

If $10<{\rm NUM}<20$    & NUM & 0.0029 & + & + & + & 93.3\% \\
and ${\rm VOL}>5$       & VOL & 0.0040 & + & + & + & \\
then \textit{malignant} &     &        &   &   &   & \\
\hline

If ${\rm TOT}>30$ & TOT & 0.0229 & - & + & + & 100.0\% \\
and ${\rm VOL}>5$ & VOL & 0.0124 & - & + & + & \\
and DEN is \textit{moderate} & DEN & 0.0325 & - & + & + & \\
then \textit{malignant} &     &        &   &   &   & \\ \hline

If VAR is \textit{marked} & VAR & 0.0044 & + & + & + & 100.0\% \\
and $10<{\rm NUM}<20$     & NUM & 0.0039 & + & + & + & \\
and IRR is \textit{moderate} & IRR  & 0.0254 & - & + & + & \\
then \textit{malignant}   &     &        &   &   &   & \\ \hline

If SIZE is \textit{moderate} & SIZE & 0.0150 & - & + & + & 92.86\% \\
and SHAPE is \textit{mild}   & SHAPE & 0.0114 & - & + & + & \\
and IRR is \textit{mild}     & IRR  & 0.0878 & - & - & + & \\
then \textit{benign}         &     &        &   &   &   & \\ \hline

\end{tabular}
\end{center}


\twocolumn

Results for them are marked as Discovery1, Discovery2 and
Discovery3. We extracted 44 statistically significant diagnostic
rules for 0.05 level of F -criterion with a conditional
probability no less than 0.75 (Discovery1). There were 30 rules
with a conditional probability no less than 0.85 (Discovery2) and
18 rules with a conditional probability no less than 0.95
(Discovery3). The most reliable 30 rules delivered a total
accuracy of 90\%, and the 18 most reliable rules performed with
96.6\% accuracy with only 3 false positive cases (3.4\%).

\section{Decision rule (model) extracted from the expert through monotone Boolean functions}

We obtained Boolean expressions for function $h \left( y_1, y_2, y_3, y_4, y_5 \right)$
(`shape and density of calcification') from the information depicted in table 1 with the
following steps:
\begin{enumerate} \item[-] Find all the maximal lower units for all chains as elementary
conjunctions; \item[-] Take the disjunction of obtained conjunctions; \item[-] Exclude the
redundant terms (conjunctions) from the end formula.
\end{enumerate}

\noindent Using 1 and 3 columns we have \[x_2 = h \left( y_1, y_2, y_3, y_4, y_5 \right) =
y_2 y_3 \vee y_2 y_4 \vee y_1 y_2 \vee y_1 y_4 \vee y_1 y_3 \vee \]
\[ \vee y_2 y_3 y_5 \vee y_2 \vee y_1 \vee y_3 y_4 y_5 \equiv y_2 \vee y_1 \vee y_3 y_4
y_5.\]

\noindent Function $g \left (w_1, w_2, w_3 \right) = w_2 \vee w_1 w_3$ we may obtain by
direct $2^{3} = 8$ questions for the expert.

\noindent Using 1 and 2 columns we have
\[f \left( \overline{x} \right) = x_2 x_3 \vee x_1 x_2 x_4 \vee x_1 x_2 \vee x_1 x_3 x_4
\vee x_1 x_3 \vee x_3 x_4 \vee x_3 \] \[ \vee x_2 x_5 \vee x_1 x_5 \vee x_4 x_5 \equiv x_1
x_2 \vee x_3 \vee \left( x_2 x_1 x_4 \right) x_5 \equiv \] \[ \left( w_2 \vee w_1w_3
\right) \left(y_1 \vee y_2 \vee y_3 y_4 y_5 \right) \vee x_3 \vee \] \[ \vee \left( y_1
\vee y_2 \vee y_3 y_4 y_5 \right) \left( w_2 \vee w_1 w_3 \right) x_4 x_5.\]


\section{Comparison of the expert model with its approximation by learning operator}

For compare rules discovered by the learning operator (Discovery
system) with the expert model we asked the expert to evaluate this
rules. Below we present some rules, discovered by Discovery
system, and radiologists comments regarding these rules as
approximation of his model.

\begin{quote}
\textsc{IF} TOTAL number of calcifications is more than 30,
\textsc{and} VOLUME is more than 5 $cm^3$,
\textsc{and} DENSITY of calcifications is moderate, \\
\textsc{THEN} Malignant.
\end{quote}

\noindent $f$-criterion significant for $0.05$. Accuracy of
diagnosis for test cases -- $100\%$. Radiologist's comment - this
rule might have promise, but I would consider it risky.

\begin{quote}
\textsc{IF} VARIATION in shape of calcifications is marked,
\textsc{and} NUMBER of calcifications is between 10 and 20,
\textsc{and} IRREGULARITY in shape of calcifications is moderate, \\
\textsc{THEN} Malignant.
\end{quote}

\noindent $f$-criterion significant for $0.05$. Accuracy of
diagnosis for test cases -- $100\%$. Radiologist's comment - I
would trust this rule.

\begin{quote}
\textsc{IF} variation in SIZE of calcifications is moderate,
\textsc{and} variation in SHAPE of calcifications is mild,
\textsc{and} IRREGULARITY in shape of calcifications is mild,
\textsc{THEN} Benign.
\end{quote}

\noindent $f$-criterion significant for $0.05$. Accuracy of
diagnosis for test cases -- $92.86\%$. Radiologist's comment - I
would trust this rule.


\section*{Acknowledgment}
This work partially supported by the Russian Science Foundation
grant 08-07-00272a and Integration projects of the Siberian
Division of the Russian Academy of science grants 47, 111, 119.



\begin{thebibliography}{99}
\bibitem{Perlovsky&Kovelerchuk-DLPC:08}
Kovalerchuk~B.\,Ya., Perlovsky~L.\,I. \emph{Dynamic logic of
phenomena and cognition}. IJCNN, 2008,
pp.~3530--3537.

\bibitem{Perlovsky} Perlovsky~L.\,I. \emph{Toward
physics of the mind: concepts, emotions, consciousness, and
symbols} // Physics of Life Reviews, 3, 2006, pp.~23--55.

\bibitem{Perlovsky2} Perlovsky~L.\,I.
\emph{Neural Networks, Fuzzy Models and Dynamic Logic} //
R.~Kohler and A.~Mehler, eds., Aspects of Automatic Text Analysis
(Festschrift in Honor of Burghard Rieger), Springer, Germany,
2007, pp.~363-386.

\bibitem{Vityaev-asian:06} Vityaev~E.\,E.
\textit{The logic of prediction} // Mathematical Logic in Asia
2005, Proceedings of the 9th Asian Logic Conference, eds.
Goncharov~S.S., Downey~R. and Ono~H., August 16--19, Novosibirsk,
Russia, World Scientific Publisher, 2006,
pp.~263--276.

\bibitem{Smerdov&Vityaev-semr:09} Smerdov~S.\,O.,
Vityaev~E.\,E. \textit{Probability, logic \& learning synthesis:
formalizing prediction concept} // Siberian Electronic
Mathematical Reports, vol.~9, 2009, pp.~340--365., in~russian,
english abstract.

\bibitem{Smerdov&Vityaev-CI:09} Vityaev~E.\,E., Smerdov~S.\,O.
\textit{New definition of prediction without logical inference}
// Proceedings of the IASTED international conference on
Computational Intelligence (CI 2009), ed. Kovalerchuk~B.\,Ya.,
August 17–-19, Honolulu, Hawaii, USA, pp.~48--54.

\bibitem{Principals} Evgenii Vityaev, Principals of brain activity, supported the
functional systems theory by P.K. Anokhin and emotional theory by P.V. Simonov,
\textit{Neiroifformatics}, v.3, N1, 2008, pp. 25-78

\bibitem{Animat} Akexander Demin, Evgenii Vityaev, Logical model of adaptive control system.
\textit{Neiroifformatics}, v.3, N1, 2008, pp. 79-107

\bibitem{Vityaev-book:06} Evgenii Vityaev. Knowledge discovery. Computational cognition.
Cognitive processes modeling. Novosibirsk State University,
Novosibirsk, 2006. pp.293.

\bibitem{Halpern:90} Halpern~J.\,Y.
\textit{An analysis of first-order logics of probability}. In:
Artificial Intelligence, 46, 1990, pp.~311--350.

\bibitem{Shiryaev-P} Shiryaev~A.\,N.
\textit{Probability}. Springer, 1995.

\bibitem{Keisler&Chang-MT:90} Keisler~H.\,J., Chang~C.\,C.
\textit{Model theory}. Elsevier, 1990.

\bibitem{Maltsev-AS:70} Maltsev~A.\,I.
\textit{Algebraic systems}. Springer-Verlag, 1973.

\bibitem{Lloyd-LP:87} Lloyd~J.W.
\textit{Foundations of logic programming}. Springer-Verlag, 1987.

\bibitem{Hansel} Hansel~G. Sur le nombre des fonctions Boolenes monotones den
variables // C. R. Acad. Sci. Paris, vol.~262, ³ 20, 1966,
pp.~1088--1090.

\bibitem{Vityaev&Kovalerchuk-finance:00} Kovalerchuk B.\,Ya., Vityaev~E.\,E.
\textit{Data mining in finance: advances in relational and hybrid
methods}. Kluwer Academic Publisher, 2000.

\bibitem{KVR-Medicine} Kovalerchuk~B.\,Ya., Vityaev~E.\,E., Ruiz~J.\,F.
\textit{Consistent and complete data and ``expert'' mining in
medicine} // Medical data mining and knowledge discovery, Springer,
2001, pp.~238--280.

\bibitem{KovTil} Kovalerchuk B, Talianski V.
Comparison of empirical and computed fuzzy values of conjunction. Fuzzy Sets and Systems
46: 49-53, 1992.

\bibitem{ProbMind} The Probabilistic Mind. Prospects for Bayesian cognitive sciense
// Eds. Nick Chater, Mike Oaksford, Oxfor University Press, 2008,
pp.536

\bibitem{ProbMod} Probabilistic models of cognition // Special issue
of the journal \textit{Trends in cognitive science}, v.10, Issue
7, 2006, pp. 287-344

\bibitem{SciDis} Scientific Discovery website. \\ http://math.nsc.ru/AP/ScientificDiscovery

\end{thebibliography}
\end{document}